# Magnetic bubbles in an M-type hexagonal ferrite observed by hollow-cone Foucault imaging and small-angle electron diffraction

Atsuhiro Kotani[1], Hiroshi Nakajima[1], Atsushi Kawaguchi[1], Yukihiro Fujibayashi[1], Kento Uchihashi[1], Keiko Shimada[2], Ken Harada[1,2], and Shigeo Mori[1,*]

[1]*Department of Materials Science, Osaka Prefecture University, Sakai, Osaka 599-8531, Japan*

[2]*Center for Emergent Matter Science (CEMS), The Institute of Physical and Chemical Research (RIKEN), Hatoyama, Saitama 350-0395, Japan*

*E-mail; mori@mtr.osakafu_u.ac.jp

ABSTRACT

We report hollow-cone imaging and small-angle electron diffraction of nanoscale magnetic textures such as magnetic-striped domains and magnetic bubbles of M-type hexagonal ferrite $BaFe_{10.35}Sc_{1.6}Mg_{0.05}O_{19}$. The advantage of the hollow-cone Foucault method is that magnetic domains with various directions of magnetization can be visualized under an infocus condition. Moreover, the contrast of magnetic domain walls in magnetic bubbles depends on the inclination angle of the illumination beam. The combination of small-angle electron diffraction and hollow-cone Foucault imaging proves that magnetization at domain walls exhibits in-plane directions in the magnetic-striped domains and magnetic bubbles.

# 1. Introduction

Lorentz transmission electron microscopy (Lorentz microscopy) is an effective observation method for visualizing nanoscale magnetic textures in real space.[1] Since its development[2], Lorentz microscopy has played a significant role in revealing interesting magnetic domains. Thus far, several methods have been employed for Lorentz microscopy including Fresnel imaging (visualization of domain walls by defocusing), Foucault imaging (visualization of domains by selecting magnetic-deflection spots)[1,3], differential phase contrast[4,5], focal-series reconstruction methods[6,7], phase microscopy (especially using hole-free phase plate)[8], and electron holography[9,10]. Further, Lorentz microscopy itself has been improved to measure magnetic domains quantitatively. For example, Foucault imaging combined with small-angle electron diffraction (SmAED) can be used to visualize magnetic domains under external magnetic fields and analyze the magnitude of magnetization.[11]–[13] Moreover, Foucault imaging using an electron biprism makes it possible to present two types of 180° magnetic domains and remove diffraction effects.[14]

Recently, we developed hollow-cone Foucault (HCF) imaging with SmAED using a conventional transmission electron microscope.[15] HCF imaging can be used to visualize several magnetic domains with various azimuthal directions of magnetization by rotating the incident beam. The advantage of HCF imaging is that both magnetic domains and domain walls are visualized simultaneously with sufficient contrast under the infocus condition. This previous study used HCF imaging revealed that 90° and 180° domains have in-plane magnetization in FeGa. However, HCF imaging has not been applied to other types of magnetic domains; further, the contrast of HCF imaging has not been clarified for other domain structures.

Therefore, in this study, we apply HCF imaging to the observation of nanoscale magnetic textures such as magnetic striped domains and magnetic bubbles in hexagonal ferrites of $BaFe_{10.35}Sc_{1.6}Mg_{0.05}O_{16}$ (BFSMO)[16]–[20]. We chose this compound for this study because it has out-of-plane magnetization in magnetic domains and in-plane magnetization in domain walls, and this type of magnetic domain has not been observed using HCF imaging. Furthermore, BFSMO exhibits magnetic bubbles, which are circularly rotating magnetic structures under external magnetic fields. Although whirling magnetic structures such as magnetic bubbles and skyrmions have attracted considerable attention because of intriguing phenomena,[21]–[29] magnetic bubbles have been observed by Fresnel imaging (out-of-focus method). Further, it is difficult to observe them using conventional Foucault imaging because they have an all in-plane direction of magnetization due to the rotating structures. In this paper, we report the observation results of HCF imaging of BFSMO and reveal the contrast reversal of magnetic bubbles that depends on the inclination angle of the illumination beam. Our results show that magnetization is oriented in the out-of-plane direction inside the magnetic bubbles, and it is oriented in the in-plane direction at the domain walls.

## 2. Experimental methods

Figure 1 shows a schematic of the configuration used for HCF imaging in this study. For magnetic domain observation, the objective lens was turned off, and then, the current in the objective mini-lens was adjusted so that a crossover (diffraction spots) was formed at the selected-area (SA) aperture plane. The SA aperture operates as an inclination angle selecting aperture in the optics of the HCF configuration. These conditions are similar to those of conventional SmAED optics.[11),30)] The electron beam irradiates the specimen with an area of approximately 85 μm in diameter with tilt angles in the X and Y directions controlled using the beam deflector system placed above the specimen. The circulating electron beam irradiates the specimen in all azimuthal directions around the optical axis. During HCF imaging, when the incident beam was rotated around the optical axis, the diffraction pattern rotated synchronously with it. For HCF imaging, some rotating spots can be selected using the SA aperture whose diameter of 20 μm corresponds to $2.60 \times 10^{-4}$ rad. The intermediate and projection lenses can be used to adjust the image magnification and camera length of diffraction patterns. The camera length used in this study was 240 m. The HCF images were obtained by integrating the images during the incident beam rotation, which enables us to visualize magnetic domains and magnetic domain walls that cause diffraction spots in various directions. The details of the optical system are described in our previous work.[15)]

This study employed two conditions of the inclination angle of the illumination beam:

$$\theta < \gamma < \theta + \beta, \quad (1)$$

$$\theta - \beta < \gamma < \theta. \quad (2)$$

In Eqs. (1) and (2), $\beta$ is the angle of the magnetic diffraction spots, $\gamma$ is the angle limited by the SA aperture, and $\theta$ is the inclination angle of the incident beam. Figure 1(b) illustrates the relationship between the angles of magnetic diffraction spots and the SA aperture, which corresponds to Eq. (1). The transmitted beam without magnetic diffraction (black spot) can pass through the SA aperture; some magnetic diffraction spots are blocked by the aperture. Hereafter, we refer to this mode as bright-field (BF) HCF imaging because the transmitted beam is used for visualization. Conversely, Fig. 1(c) shows the situation corresponding to Eq. 2. In this condition, the SA aperture blocks the transmitted beam, and thus, it is referred to as dark-field (DF) HCF imaging. Note that the beam-convergence semi-angle α should be smaller than β to observe magnetic diffraction spots.

HCF imaging was performed using a transmission electron microscope JEM-2100F (200 kV, JEOL Co. Ltd., Japan). An external magnetic field was applied using the objective lens. A single crystal of BFSMO was grown via the floating zone method.[18)] The *c* plane, which was oriented perpendicular to the magnetic easy axis, was polished and then thinned by Ar-ion milling.

## 3. Results and discussion

First, we clarify the relationship between the Fresnel image and SmAED pattern of BFSMO (Fig. 2) for HCF imaging. The contrast of the Fresnel image indicates that magnetization is oriented in the out-of-plane direction in the magnetic domains and in the in-plane direction in the magnetic domain walls, as reported in Ref [18)]. The SmAED pattern recorded with a 240 m camera length shows a transmitted spot 000 and magnetic diffraction spots. Magnetic diffraction spots are formed in a ring shape because the magnetic domains show a maze pattern with all azimuthal directions. The period calculated from these magnetic spots (5.84 μrad) is 430 nm, which corresponds to twice the magnetic-domain width. As indicated by the red arrows in the inset of Fig. 2(a), the domain walls have alternating magnetizations in the upwards and downwards directions and a magnetic period of approximately 440 nm. This agrees with the periodic arrangement of the SmAED pattern. This indicates that the magnetic diffraction spots originate from the domain walls, as demonstrated in the previous study[31)].

To perform the HCF imaging of magnetic bubbles, we applied an external magnetic field of 2 T and then reduced it to zero. This process changed the striped domains into magnetic bubbles. Figure 3(a) shows a Fresnel image in which magnetic stripes and magnetic bubbles coexist in a remanent magnetization state. The red arrows of Fig. 3(b) indicate the directions of magnetization inferred from the contrast of the Fresnel image. The Fresnel image demonstrates that the in-plane direction component of the magnetization rotates clockwise or counterclockwise in the domain walls of the magnetic bubbles. Similar to Fig. 2, the SmAED pattern depicts ring-shaped scattering because of the various azimuthal directions at the domain walls; again, diffraction spots corresponding to the periodic arrangement of domain walls can be seen clearly.

The HCF images of magnetic striped domains and magnetic bubbles are shown in Figs. 3(c) and 3(d). Figures 3(c) and 3(d) were obtained under conditions corresponding to Eqs. (1) and (2), respectively: $\beta$ = 5.84 μrad, $\gamma$ = 2.60 × 10$^{-4}$ rad, and $\theta$ = 15.3 μrad for Eq. (1) and $\theta$ = 18.8 μrad for Eq. (2). In Fig. 3(c), the transmitted spot 000 was selected in addition to some magnetic diffraction spots with the SA aperture (Eq. (1)). In this condition, the domains are depicted as bright areas, while the domain walls are dark lines, as shown in Fig. 3(c). The contrast can be explained as follows: As demonstrated in Fig. 2, magnetic diffraction spots in the SmAED pattern originate from the domain walls, which have in-plane magnetization. The transmitted spot 000 was caused by the vacuum or domains in the specimen with the magnetization parallel to the incident beam. Thus, the domain walls show as dark lines, compared with the domains themselves, when some of the periodic spots are blocked by the aperture during the hollow cone irradiation process.

Conversely, Fig. 3(d) shows the HCF image captured using some of the magnetic diffraction spots and by excluding the transmitted spot 000 [Eq. (2)]. Unlike Fig. 3(c), the magnetic domain walls appear as bright lines under these experimental conditions. The contrast change is attributed to the domain walls causing magnetic spots that were selected with the aperture.

The contrast reversal is clearly confirmed in Fig. 4. These images are magnified images of BF and DF HCF images around magnetic bubbles. The circular domain walls are visualized as dark and bright lines in Fig. 4(a) and 4(b), respectively. The intensity profiles along the line X–Y indicate contrast reversal at the domain wall. As explained above, the regions with the in-plane magnetization component can be seen as bright in the magnetic bubbles because the DF HCF image was formed only using magnetic spots.

Magnetic bubbles comprise domain walls aligned with all azimuthal directions. Thus, conventional Foucault imaging cannot visualize magnetic bubbles as it can only depict domains with one particular direction of magnetization. HCF imaging, which has the ability to visualize domains with directional scattering in multiple directions, thus supports magnetic domain observation. In addition, Fresnel images (e.g., Fig. 3b) indicate domain walls as a pair of bright and dark lines. HCF imaging can visualize magnetic domain walls as a single bright or dark line (BF image of Fig. 4a and DF image of Fig. 4b). Consequently, by combination with Fresnel imaging, HCF imaging makes it possible to understand the presence of domain walls even when bend contours and Fresnel fringes from twin boundaries and edges exist.

These experimental results can be understood in a similar manner as those in a previous study of 180° magnetic domains in FeGa[15]. Noticeably, the relationship between domain and domain wall contrast in this study is opposite to that of FeGa. In the 180° domain structure of FeGa, the magnetization is oriented in the in-plane direction inside the domains, and it is oriented in the out-of-plane direction in the domain wall. As a result, the domain walls were observed as bright lines in a BF HCF image and as dark lines in a DF HCF image. Considering that magnetic spots were caused by in-plane magnetic components, the contrast of the previous study can be explained in a similar manner to that of this study[15]. In addition, phase microscopy studies support a magnetic domain structure similar to that obtained by HCF imaging.[7,8] Therefore, the present HCF imaging experimental results demonstrate that magnetization at the domain walls points to the in-plane direction in BFSMO.

## Conclusions

Magnetic textures and the magnetization distributions of hexagonal ferrite BFSMO were investigated using SmAED and HCF imaging. We demonstrated that HCF imaging can visualize magnetic domain walls and magnetic bubbles; the contrast depends on the inclination angle of the illumination beam. In BFSMO, BF HCF imaging depicted domain walls as dark lines, whereas in DF HCF imaging, they were depicted as bright lines. HCF imaging combined with SmAED demonstrated that this material exhibits in-plane magnetization in the magnetic domain walls.

## Acknowledgments

This work was supported by KAKENHI, Grant-in-Aid for Scientific Research (B) JP18H03475 and (S) JP19H05625. This work was also in part supported by JSPS KAKENHI Grant Number JP19H05814 (Grant-in-Aid for Scientific Research on Innovative Areas "Interface Ionics").

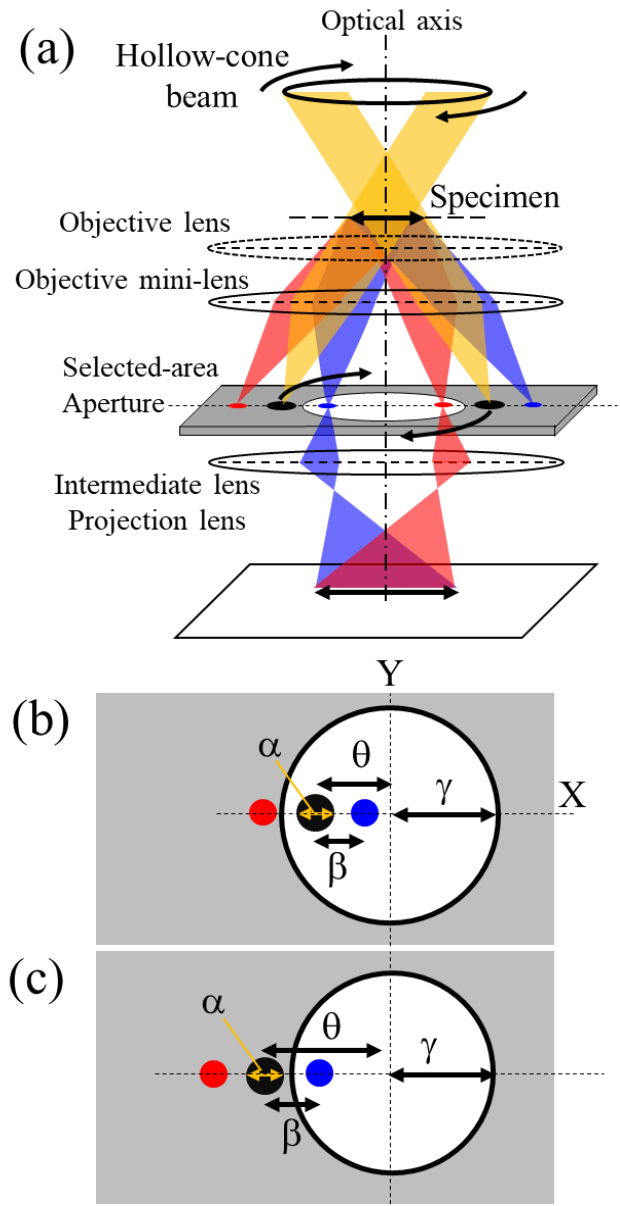

Fig. 1. (a) Schematic of optical configuration of hollow-cone Foucault imaging. Yellow and red (blue) lines represent the direct and diffracted beams, respectively. Diffraction spots with all azimuthal directions such as the red and blue lines can pass through the selected-area aperture to appear in the hollow-cone Foucault image. (b) Relationship between diffraction spots and SA aperture in a bright-field mode. (c) Relationship between diffraction spots and SA aperture in a dark-field mode. The labels indicate the inclination angle of the illumination beam θ, diffraction angle β, and angle limited by the selected-area aperture γ. The angle α represents the divergence semi-angle. (b) and (c) correspond to the conditions of Eqs. (1) and (2), respectively.

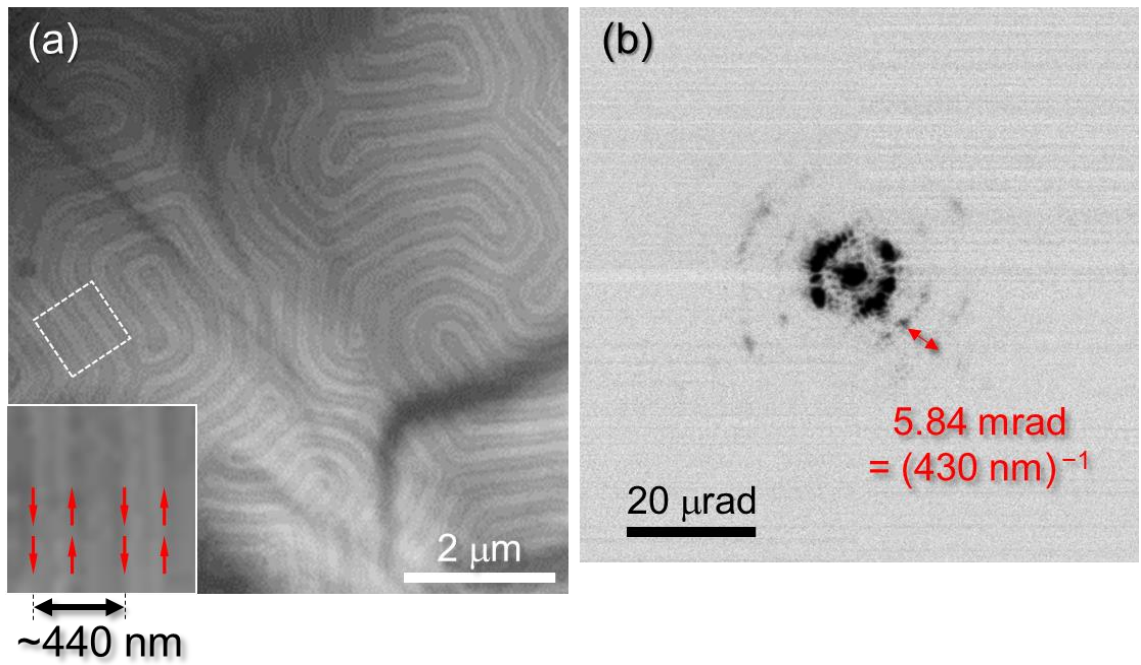

Fig. 2. (a) Fresnel image and (b) small-angle electron diffraction (SmAED) pattern of magnetic striped domains in BFSMO. In the inset of (a), the region indicated by the white dotted square is magnified. Red arrows represent the directions of the magnetization estimated from the Fresnel image.

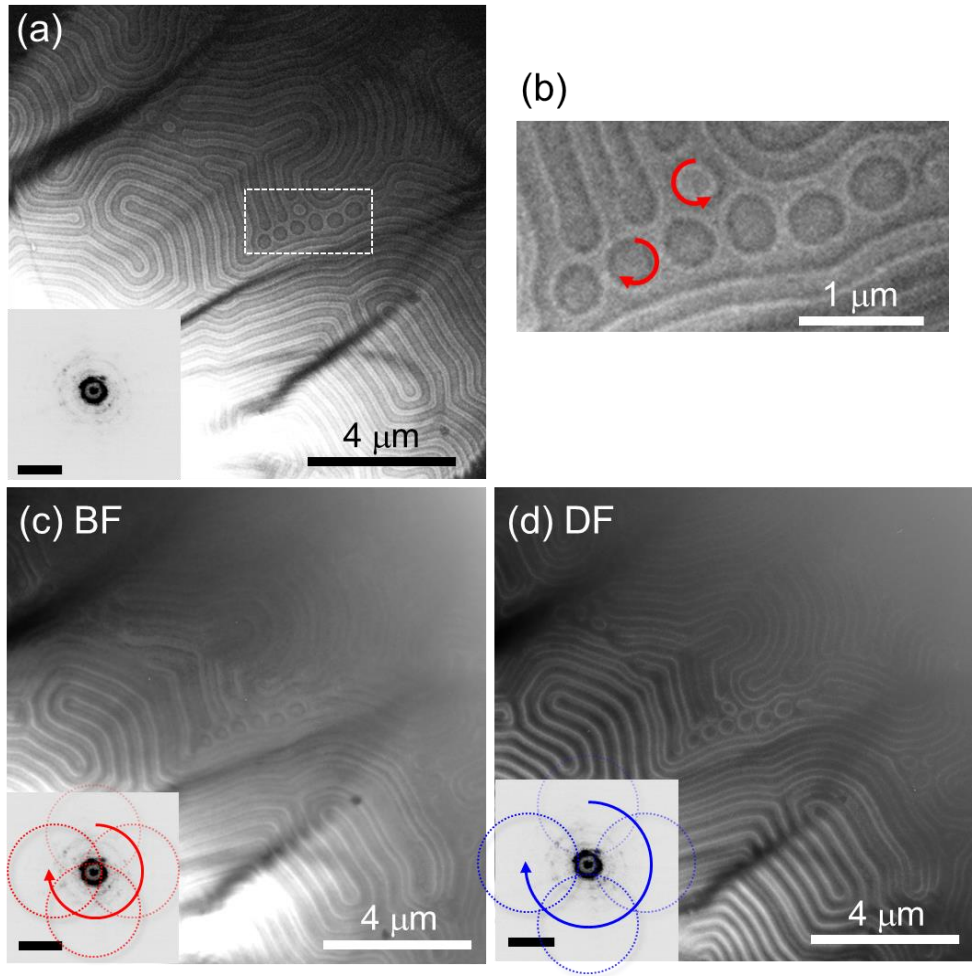

Fig. 3. (a) Fresnel image of a coexisting state of striped domains and magnetic bubbles at room temperature under no magnetic field after applying an external field of 2 T. The inset is a small-angle electron diffraction pattern (scale bar 20 μrad). (b) Magnified image of the area surrounded by the white dotted rectangle in (a). Red arrows indicate the directions of magnetization. (c) Bright-field and (d) dark-field HCF images. The insets of (c) and (d) show the ranges selected with the selected-area aperture for the HCF imaging (scale bar 20 μrad).

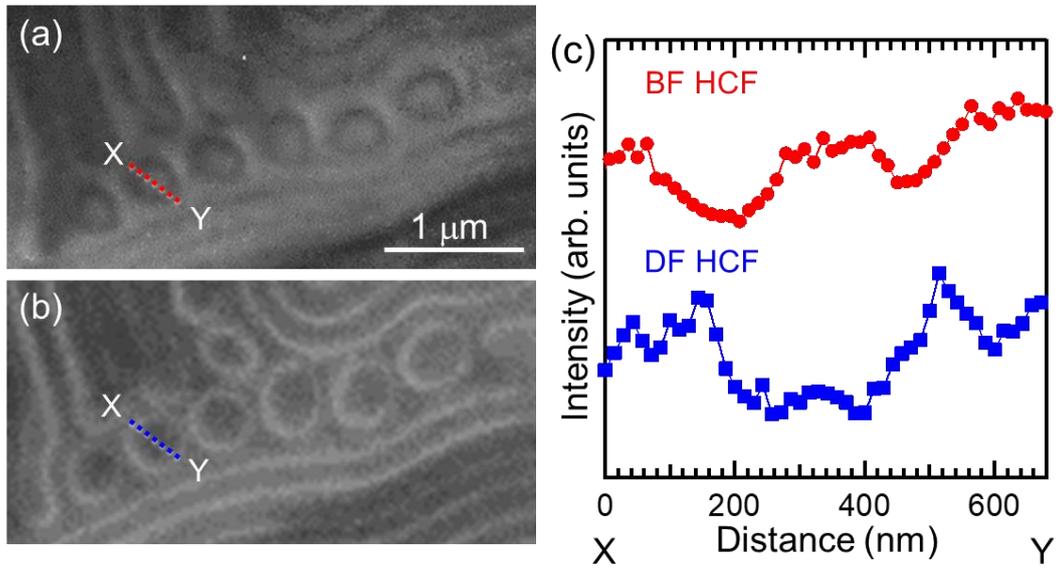

Fig. 4. (a) Bright-field (BF) HCF image of BFSMO, (b) dark-field (DF) HCF image, and (c) intensity profiles along the red and blue lines (X–Y) in panels (a) and (b). The field of view in panels (a) and (b) is the same as that of Fig. 3(b). The contrast of the magnetic bubbles in the BF and DF HCF images are opposite to each other.